\documentclass[12pt]{article}

\newread\testifexists
\def\GetIfExists #1 {\immediate\openin\testifexists=#1
    \ifeof\testifexists\immediate\closein\testifexists\else
    \immediate\closein\testifexists\input #1\fi}

\usepackage{gthstyle}\usepackage{amsfonts}
\usepackage{amssymb}
\immediate\openin\testifexists=e
    \ifeof\testifexists\immediate\closein\testifexists\else
    \immediate\closein\testifexists\input e\fipsf
\def\Bbb#1{\setbox0=\hbox{$\tt #1$}  \copy0\kern-\wd0\kern .1em\copy0}
\def\bbf#1{\setbox0=\hbox{$#1$} \kern-.025em\copy0\kern-\wd0
        \kern.05em\copy0\kern-\wd0 \kern-.025em\raise.0433em\box0}
     \def\s{\sigma}

      \def\W{\Omega}

\def\ffract#1#2{\raise .3 em\hbox{$\scriptstyle#1$}\kern-.25em/
                \kern-.2em\lower .2 em \hbox{$\scriptstyle#2$}}

\def\part#1#2{{\partial#1\over\partial#2}}

\newcommand{\be}{\begin{eqnarray}}
\newcommand{\ee}{\end{eqnarray}}

\newcommand{\bi}[1]{\begin{itemize}\item[#1]}

\newcommand{\ei}{\end{itemize}}

\newcommand{\newsec}[1]{\section{#1}\setcounter{equation}{0}}
\def\printversion{\setlength{\textheight}{9in}\setlength{\oddsidemargin}{0in}
    \setlength{\textwidth}{6.3in}\setlength{\topmargin}{-0.1in}}

\newcommand {\eel}[1]{\label{#1}\end{eqnarray}} % equationnumbers

\printversion 
\begin{document} \begin{titlepage}

\title{\normalsize \hfill ITP-UU-07/4 \\ \hfill SPIN-07/4 \\ \hfill {\tt quant-ph/0701097}\\ \vskip 8mm \Large\bf ON
THE FREE-WILL POSTULATE IN QUANTUM MECHANICS}
 
\author{Gerard 't~Hooft}
\date{\normalsize Institute for Theoretical Physics \\
Utrecht University \\ and
\medskip \\ Spinoza Institute \\ Postbox 80.195 \\ 3508 TD
Utrecht, the Netherlands \smallskip \\ e-mail: \tt
g.thooft@phys.uu.nl \\ internet: \tt
http://www.phys.uu.nl/\~{}thooft/}

\maketitle

\begin{quotation} \noindent {\large\bf Abstract } \medskip \\
The so-called ``free will axiom" is an essential ingredient in many discussions concerning hidden variables in quantum
mechanics. In this paper we argue that ``free will'' can be defined in different ways. The definition usually employed
is clearly invalid in strictly deterministic theories. A different, more precise formulation is proposed here, defining
a condition that may well be a more suitable one to impose on theoretical constructions and models. Our axiom, to be
referred to as the `unconstrained initial state' condition, has consequences similar to ``free will", but does not
clash with determinism, and appears to lead to different conclusions concerning causality and locality in quantum
mechanics. Models proposed earlier by this author fall in this category. Imposing our `unconstrained initial state'
condition on a deterministic theory underlying Quantum Mechanics, appears to lead to a \emph{restricted} free will
condition in the quantum system: an observer has the free will to modify the setting of a measuring device, but has no
control over the phase of its wave function. The dismissal of the usual ``free will" concept does not have any
consequences for our views and interpretations of human activities in daily life, and the way our minds function, but
it requires a more careful discussion on what, in practice, free will actually amounts to.
\end{quotation}

%\vfill \flushleft{\today}

\end{titlepage}

\newsec{General Introduction}

There may be different motivations for studying the ontological nature of quantum mechanics. Three different starting
points appear to stand out. The first is a fundamental unease with the way quantum mechanics is presented in standard
text books. It appears that quantum mechanics gives a description of features that cannot be directly observed, called
``wave functions", in order to enable us to understand and predict outcomes of observations and measurements. The
predictions are often ``uncertain", which means that various possible outcomes of measurements are predicted, and only
the statistical distributions for the outcomes can be predicted with certainty, when experiments are repeated many
times. One could side with Einstein when he argued that this is an unsatisfactory state of affairs for any theory, and
assert that only theories that, under ideal circumstances, would predict single events with certainty should be
considered acceptable. The fact that, in practice, neither the initial state nor the infinitely precise values of
constants of nature will be exactly known, should be the only acceptable source of uncertainties in the predictions.

Many researchers express other reasons for being unhappy with the standard formulation of Quantum Mechanics, which is
that not only microscopic events, but also the acts of observation, storage and the interpretation of these
observations must, in principle, be described by Quantum Mechanics. There must be some point at which observations turn
into classical certainties, which is described as the ``collapse of the wave function". When and how does such a
``collapse" take place? This transition point between quantum mechanical and classical, between wave function and
probability function, is causing concern.

The third group of investigators is asking more pragmatic questions. The Standard Model of the elementary particles
describes in meticulous detail how all known constituents of matter behave. It is an entirely quantum mechanical
description, and one can say that Quantum Mechanics has emerged as an absolutely indispensable framework for this
description. Yet the model fails when gravitational forces come into play. Now, even though there is a strongly
advocated approach to this problem called ``Superstring theory", boasting to be the one and only way to address the
problem of quantizing gravity completely successfully, one cannot help observing that this theory is still lacking any
kind of strong foundations in rigorous logic. This theory hangs together with asymptotic perturbative expansions, where
different domains appear to be related by duality relations, but in the cases of essential interest, these expansions
tend to diverge, and their true basis in any kind of rigorous formulation is still obscure. The advocates of
superstring theory appear to be confident that these problems will be resolved, but one cannot help suspecting that the
present formulation of the theory does not appear to give us hints as to where such solutions will come from.

An essential element of superstring theory is the assertion that it should be an \emph{all-embracing theory}, a
``theory of everything". If so, then it should also, in a natural manner, answer some of the questions concerning
quantum mechanics itself. What is Quantum Mechanics? Should a ``theory of everything" not also explain (at least in
principle) how all those fluctuations in space and time could have emerged, what their origin might be, in terms of
rigorous equations of motion, how exactly galaxies, stars and planets came into being? Quantum Mechanics as we know it
will never give answers to such questions, since only statistical distributions will be predicted. Is this not a
shortcoming for a purported ``theory of everything? It is this author's opinion that the most likely place to search
for new \emph{conceptual} solutions, solutions that not only answer the question what the non-perturbative nature of
superstring theory is, but also address all those other related problems, may come from asking a next generation of
questions. In the past, one such question has been the incorporation of black holes into string theories. This turned
out to have been a fruitful exercise, leading to many new insights. It is this author's belief that similar questions
will have to be asked concerning the deeper nature of Quantum Mechanics, and that attempts to answer such questions
will again lead to further insights. Superstring theorists thus-far have turned their backs to the issues raised by
those who try to understand Quantum Mechanics at a deeper level. It is time to face these questions.

\newsec{Free Will}

A class of very important questions arose when John Bell formulated his famous in\-equal\-ities\cite{JSB}. Indeed, when
one attempts to construct models that visualize what might be going on in a quantum mechanical process, one finds that
deterministic interpretations usually lead to predictions that would obey his inequalities, while it is well understood
that quantum mechanical predictions violate them. In attempts to get into grips with this situation, and to derive its
consequences for deterministic theories, the concept of ``free will" was introduced. Basically, it assumes that any
`observer' has the freedom, at all times and all places, to choose, at will, what variables to observe and measure.
Clashes with Bell's inequalities arise as soon as the observer is allowed to choose between sets of observables that
are mutually non commuting.

To change the set-up of an experiment, say in order to measure the \(x\)-component rather than the \(z\)-component of
the spin of a particle, requires a macroscopic modification of the apparatus, but the researchers see no reason why
such a modification should affect the wave function of an approaching particle. That, purportedly, would imply a
non-local interaction of a type that one would wish to preclude.

This way of thinking is a natural continuation of the argumentation process that has been extremely successful in the
conceptual construction of twentieth century theories of physics, such as Special and General Relativity, as well as
Quantum Mechanics itself: rather than directly discussing the microscopic physical variables, we have learned to talk
in terms of `observers', `measurements' and spaces of possible outcomes of measurements, while avoiding the microscopic
variables themselves as long as is possible, just because our human experiences in the macroscopic world might be
misleading us. In this paper, we wish to point out the danger of this kind of argumentation. Many researchers are led
to believe that the microscopic world is controlled by `a different kind of logic' than our classical logic. We insist
that there exists only one kind of logic, even if the observed phenomena are difficult to interpret.

Just imagine that we would be living in a \emph{completely deterministic world}. Would the notion of `free will' that
is usually employed be correct?

My answer is: of course not! A lengthy discussion will now be given, but it should be emphasized that our basic
observation is an extremely simple one, falling in the category of the little boy shouting that the Emperor is not
wearing any clothes.

Can we, at will, modify the set-up of an apparatus without modifying any of the particles that we are measuring? Can we
change our initial intention to measure the \(z\)-component of a spin, into measuring its \(x\)-component, without at
the same time affecting the delicate, \emph{microscopic} variables in our vicinity, and in particular the particle we
intend to investigate? Suppose that we replace our machines, including our own decision making processes, by truly
macroscopic devices such as planets moving in their Newtonian orbits. Can the Planet Mercury, while `detecting' the
planet Pluto, decide `at will' to go sit somewhere else in its orbit around the Sun, while Pluto continues in an
unperturbed way? Of course not. If Mercury had to be imagined somewhere else in its orbit, the entire past of the
planetary system would have to be modified, including the way Pluto moves, even if it sits somewhere else in the Solar
System. In short, one cannot modify the present without also modifying the past (and, of course, the future). Can one
modify the past in such a way that, at the present, \emph{only} the setting of our measuring device are modified, but
not the particles and spins we wish to measure? Can we modify our settings without modifying the \emph{wave functions}
of the particles approaching us? This is the question we address, in particular in Section \ref{Coffee.sec}. `Free
will', meaning a modification of our actions without corresponding modifications of our past, is impossible. In chapter
\ref{Coffee.sec}, we will further address the practical consequences of this `absence of free will', which we claim to
be much less drastic than often implicated.

Most authors appear not to be ready to question the nature of their `free will' assumption. The following citations
from recent literature could serve to demonstrate this: in ref\cite{BG}, Bassi and Ghirardi state: ``Needless to say,
the \textbf{FREE} axiom [the free-will assumption, a.] must be true, thus \(B\) is free to measure along any triple of
directions. ...". But it is not further explained what `free' here means.

In Ref.~\cite{KochenConway}, Conway and Kochen assume `free will' to be `just that the experimenter can freely choose
to make any one of a small number of observations'. Remarkably, they continue to say that the fact they claim to
derive, this ``failure to predict" is `a merit rather than a defect, since these results involve free decisions that
the universe has not yet made'. Again, no further explanation of what `free' here means is considered necessary.

In Ref\cite{RT}, R.~Tumulka attempts to``point out three flaws in Conway and Kochen's argument", but the assumption
itself, that of the existence of free will, is not among the purported flaws: ``As a consequence of (1)," he notes,
``we have to abandon one of the four incompatible premises. It seems to me that any theory violating the freedom
assumption invokes a conspiracy and should be regarded as unsatisfactory", and later: ``We should require a physical
theory to be non-conspirational, which means here that it can cope with arbitrary choices of the experimenters, as if
they had free will (no matter whether or not there exists ``genuine" free will). A theory seems unsatisfactory if
somehow the initial conditions of the universe are so contrived that EPR pairs always know in advance which magnetic
fields the experimenters will choose." `Conspirational' could mean the modification in the past that is required to
modify the present. In a deterministic theory, this `conspiration' is diffucult to object against, so presumably what
is meant here is that Nature conspired to adapt the \emph{wave functions} as well to the new situation. The point we
wish to make is, that in a deterministic theory there are no wave functions, in particular no phases of wave functions;
the phases we use to describe them are artifacts of our calculational procedures, and they could well be determined by
what happened in the past. As long as this is not exactly understood, there appears to be no contradiction.

In Ref\cite{NewScientist}, the mathematician Conway, when interviewed, exclaims: ``We have to believe in free will to
do anything; I Believe I am free to drink this cup of coffee, or throw it across the room. I believe I am free in
choosing to have this conversation." But of course Conway should know that, in spite of his apparent freedom to throw
his coffee across the room or not, whatever he actually does \emph{is} determined by laws of physics, not be some
mysterious, unspecified, `free will'. This, at least, is the real implication of the assumption of determinism. He is
free to choose what to do, but this does not mean that his decision would have no roots in the past.

So much for the literature. (See Sect.~\ref{Coffee.sec}). We dismiss all unquestioned `free will' assumptions in
physics as being not worthy of a mathematically rigorous theory.

At this point, however, the reader may ask, yes, but don't we introduce `operators' all the time in Quantum Mechanics?
We apply the operator \(\hat p_i\), generating an infinitesimal displacement, the operators \(\hat a_i\) and \(\hat
a_i^\dagger\) that annihilate or create some particle in a given quantum state; is this not the same as an `at will'
modification of the wave functions? Surely it is, and this is exactly the point this author wishes to make. Such
operations have become commonplace in the way we think about Quantum mechanics. They are extremely useful in
calculating something, but we should not forget what we are really doing. The situation becomes a bit more clear if we
adopt the \emph{Heisenberg representation} rather than the Scr\"odinger representation of quantum states. In the
Heisenberg representation, we can calculate how operators \(\W(t)\) behave as a function of time. Suppose we let an
operator \(a_i(\vec x,\,t)\) act on a state, which means, more or less, that we remove a particle at the point \(\vec
x\), at time \(t\). A different state is then obtained, \emph{in which both the future and the past development of
operators look different from what they were in the old state}! A state can only be modified if both its past and its
future are modified as well. Only such modifications lead to states that obey the same physical laws as before. Again,
we stress that there is nothing very deep in this observation --- it is so obvious that one might tend to overlook it.
We should acknowledge the simple fact that if we have any free will at all, a modification of our actions will modify
our past as well as our future. We see that this even holds for Quantum Mechanics; it certainly holds for deterministic
systems.

But then there are more questions to be asked; why would one want to introduce `free will' at all? What is so important
about it? Indeed, an assumption of this nature \emph{is} needed to formulate realistic theories. Without it one cannot
build any model of physics at all! This is explained next.

\newsec{The unconstrained initial state}

In this paper, it is advocated to drop the old way of thinking entirely: stop arguing about `observers', `systems',
`apparatus' and `measuring results'. Instead, let us concentrate on `physical variables', `physical states', and
`complete descriptions' of these physical states. These are the true ingredients if we wish to build a model. One could
have the Standard Model of the elementary particles in mind, or some string theory, or the planets in their orbits
around the Sun. All we should demand is that the model in question obeys the most rigid requirements of internal logic.
Our model should consist of a \emph{complete description} of its physical variables, the values they can take, and the
laws they obey while evolving. The notion of time has to be introduced if only to distinguish \emph{cause} from
\emph{effect}: cause must always precede effect. If we would not have such a notion of time, we would not know in which
order the `laws of nature' that we might have postulated, should be applied. Since laws of nature tend to generate
extremely complex behavior, their effects will surely depend on the order at which they are applied, and our notion of
time will establish that order.

In most other respects, our model of the world could be almost anything. But let us momentarily assume that it is
deterministic in the classical sense. Let us accept what Omar Khayamm said, which is that the events in the entire
Universe were essentially fixed at the same moment it came into being: the conditions at the Big Bang were decisive all
the way up to and including the events characterizing the `final apocalypse'. Where does the notion of `free will'
appear?

Surprisingly, perhaps, we claim now that free will is everywhere, but it is not the notion that was assumed as an
`axiom' by today's quantum scientists. We see the situation as follows. In Nature, the Laws determining its evolution
are complex. This means that, in the vast majority of cases, one will have no way to foresee exactly what will happen.
Only after meticulously painstaking calculations, from beginning to end, one might be able to look forward a bit, but
very soon, one will be forced to make crude approximations. The true values of Nature's degrees of freedom will not be
known for sure --- one will have to make `educated guesses'. Conversely, if we wish to understand why and how a certain
situation in this Universe has arisen, we have to make numerous guesses concerning the past, and eventually select the
one that fits best with everything we know. In our model, we will \emph{only} be able to perform such tasks if we
possess some notion of the \emph{complete class} of all possible configurations of our variables. For \emph{every
member} of this class, our model should produce reasonable predictions. Even if, in the real world, only very limited
subsets of all possibilities will ever be realized anywhere at any time, our model must be able to describe \emph{all}
eventualities.

If we would have been deprived of the possibility to freely choose our initial states, we would never be able to rely
on our model; we would not know whether our model makes sense at all. In short, we must demand that our model gives
credible scenarios for a universe \emph{for any choice of the initial conditions}!

This is the free will axiom in its modified form. This, we claim, is why one should really want `free will' to be
there. It is not the free will to modify the present without affecting the past, but it is the freedom to choose the
initial state, regardless its past, to check what would happen in the future.

Indeed, when Tumulka, in the quoted text, talks about conspiracy, stating that conspiratorial theories appear to be
unacceptable, it was actually this modified form of free will that he had in mind. \textit{But this is not the free
will that is assumed in the Conway-Kochen argument!}

One cannot modify the present without assuming some modification of the past. Indeed, the modification of the past that
would be associated with a tiny change in the present must have been quite complex, and almost certainly it affects
particles whose spin one is about to measure. We return to this in Section \ref{Coffee.sec}.

\newsec{Operators in Quantum Mechanics}\label{QuOp.sec} With regard to determinism, the
most conspicuous feature of Quantum Mechanics is the central position of operators. If an operator refers to an
observable, it is assumed to describe `reality' in some form. We call it a `beable' then. But an operator of the same
form, under different circumstances, may refer to a replacement. It is then called a `changeable'. For instance, the
operator \(\s_3\) for an electron may be a beable, when used to measure its spin, if the spin is oriented in the
\(z\)-direction, but the \emph{same} operator may be used to switch the spin, if we refer to spin in the
\(x\)-direction or the \(y\)-direction; then, it is a changeable.

Our point is, that by itself this does not turn Quantum Mechanics into a non determin\-istic theory. We could also
introduce such operators in Newtonian gravity, for instance an operator \(M_{12}\) that replaces the planet Mercury
from position 1 (if it happens to be there) to position 2, somewhere else in the solar system. In Newtonian gravity,
operators of this sort do not seem to be very useful; in Quantum Mechanics, they are. Indeed, only when using these
operators, one can establish that the theory is invariant under rotations. At least one classical model exists in which
one can enhance the symmetry by referring to quantum operators, even though the theory itself is deterministic, see
Ref.~\cite{GtHCAsymm}. There are also examples where the methods of Quantum Field Theory are used to solve a classical
theory, such as the two-dimensional Ising Model\cite{Kaufman}.

In a deterministic theory, beables and changables are two distinct varieties of operators. The first an observable
quantity, in the deterministic sense. A beable does not affect the ontological status of a system, and therefore, by
fiat, all beables commute with one another at all times. Operators not commuting with one or more beables are called
changeables. When, out of some purported `free will', an observer changes his mind as to what to measure, by modifying
the setting of his or her apparatus, this could be seen as the application of a changeable. According to both Quantum
Mechanics and any deterministic theory, applying a changeable now, at time \(t=0\), implies it is applied at all times,
before and after \(t=0\): operators evolve.

\begin{figure}[h]\setcounter{figure}{0} \begin{quotation}
 \epsfxsize=120 mm\epsfbox{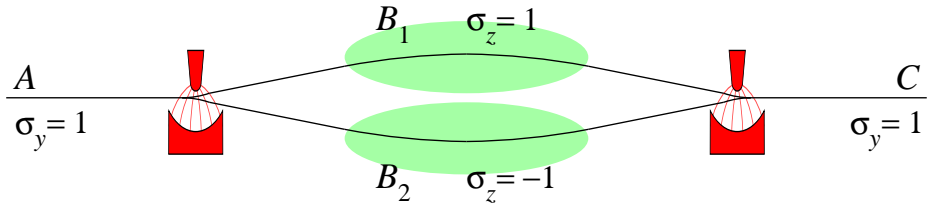}
  \caption{\footnotesize  hypothetical experiment with two Stern-Gerlach beam splitters. The initial spin is in the
  \(y\) direction; the inhomogeneous magnetic fields splits the beam into one with \(\s_x=1\) and one with
  \(\s_x=-1\). The ovals are magnetic fields turning the beams back. After the beams reunite, the spin is again 1
  in the \(y\) direction.}
  \label{figure01.fig}\end{quotation}
\end{figure}

In Figure \ref{figure01.fig}, a conceivable experiment with electrons is described. The point is not whether the
experiment can actually be performed or not. In principle, it is possible. The question is how to describe it in a
deterministic formalism. It would be a mistake to think that, at the points \(A\) and \(C\), \(\s_y\) is the beable,
and at the points \(B_{1,2}\) it would be \(\s_z\). Within Quantum Mechanics, it is clear how to describe the time
evolution of these operators. If the electrons are allowed to separate and rejoin unperturbed, the beables during the
moment of separation are quite complex, and they seem to be non-local. However, the usual arguments claiming that they
\emph{have} to be non-local, do not apply in our approach to determinism. Whatever the local beables are, they cannot
be expressed in terms of \(\s_y\) or \(\s_z\).

\newsec{Conway's coffee, conspiracies and quantum phases.}\label{Coffee.sec}

The need for an `absolute' free will is understandable. Could there exist any `conspiracy' to prevent Conway to throw
his coffee across the room during his interview? Of course, no such conspiracy is needed, but the assumption that his
decision to do so depends on microscopic events in the past, even the distant past, is quite reasonable in any
deterministic theory, even though, in no way can his actions be foreseen. It is easy to argue that, even the best
conceivable computer, cannot compute ahead of time what Mr.~Conway will do, simply because Nature does her own
calculations much faster than any man-made construction, made out of parts existing in Nature, can ever do. There is no
need to demand for more free will than that.

The unconstrained initial state condition states that, regardless what Conway does with his coffee, the deterministic
equations should determine what will happen next. However, these deterministic equations cannot contain wave functions,
and \emph{therefore}, one cannot demand to have the free will to modify the settings of a detector (or throw the coffee
in some direction), \emph{without even affecting the wave functions} of the objects measured. According to our
deterministic theories, these wave functions are man-made artifacts, and they can therefore not be kept unaffected.
These wave functions may well depend critically on past events, in a fully conspiratorial manner.

A quite similar situation is known to exist in local gauge theories for elementary particles. Fixing the gauge by some
gauge condition may generate field configurations that depend in a conspiratorial way on the past or the future, but
this has no effect on the physically observable events, just because these are gauge-independent. In the same vein, the
dependence on wave functions may appear to be conspiratorial, just because the wave functions as such are unobservable.

\end{document}